\documentclass[letterpaper]{sig-alternate}
\usepackage{amsfonts}
\usepackage{url}
\usepackage{theorem}
\usepackage{floatflt}
\usepackage{graphics}
\usepackage{epsfig}
\usepackage[T1]{fontenc}
\usepackage{amsmath}

\newcommand{\ignore}[1]{}

\theoremstyle{plain} 
\conferenceinfo{WPES'03,}{October 30, 2003, Washington, DC, USA.}
\CopyrightYear{2003}
\crdata{1-58113-776-1/03/0010}
\title{New Covert Channels in HTTP} 
\subtitle{Adding Unwitting Web Browsers to Anonymity Sets}
\numberofauthors{1}
\author{
\alignauthor Matthias Bauer\\
\affaddr{Institut f\"ur Informatik}\\
\affaddr{Martensstrasse 3}\\
\affaddr{91058 Erlangen, Germany}\\
\email{matthiasb@acm.org}
}

\begin{document}
\maketitle

\begin{abstract}
This paper presents new methods enabling anonymous communication
on the Internet. We describe a new protocol that allows us to
create an anonymous overlay network by exploiting the web browsing
activities of regular users.  We show that the overlay network
provides an anonymity set greater than the set of senders and receivers
in a realistic threat model. In particular, the protocol provides
unobservability in our threat model.
\end{abstract}


\section{Introduction}
Privacy on the Internet gains importance as most network
activity can be linked to a user's identity. Proposed solutions 
that use Chaumian mixes show certain traffic patterns
if not every user runs a node in the system.  
We describe a realistic threat model and present
a new set of protocols that allow unobservable communication.

In 1981, David Chaum presented the concept of {\em mixes},
a protocol to provide sender--receiver {\em unlinkability} under
standard cryptographic assumptions.  Unlinkability
means that an observer does not learn anything to improve
her guesses on who communicates with whom (The {\em
a--priori} probability of two entities being related is
equal to the {\em a--posteriori} probability).
The notion of the {\em anonymity set} is essential when
measuring anonymity. It is the set of all possible subjects
who might cause an action \cite{Koehn}. In the context of classical
communication protocols, it consists of the senders and
receivers.

Currently employed implementations
of mix networks are, for example, the Cypherpunks \cite{remailer-history} and
Mixmaster \cite{mixmaster-spec} remailers and the nascent
Mixminion project \cite{mixminion}.
Chaum's work motivated other schemes
which avoid expensive decryption at each step, to minimize
delay, for example, Crowds \cite{Crowds} and 
Onion Routing \cite{Syverson}.

In practice, most users of these systems do not run mix nodes
themselves. They cause traffic patterns  to and from the
set of mix nodes, which a global, passive adversary can use to
reduce the anonymity provided by the systems.
Possible attacks by such an observer include intersection,
timing and packet counting attacks on
remailers and other systems
derived from Chaumian mixes \cite{Loki2, back01, Raymond01}.
Suggested solutions introduce {\em cover traffic}
into the protocols. This is achieved mainly by having the senders
inject {\em dummy messages} \cite{langos02}, which are discarded
at some mix. 

{\em Unobservability} is a stronger property than unlinkability,
meaning that an observer cannot tell if messages are being sent or
received {\em at all}.

This paper presents techniques to
enlarge the anonymity set by including noninvolved subjects who provide cover traffic
for the protocol in question.
Our approach is to hide communication within
transit traffic going through HTTP browsers.

In our model, the adversary cannot
distinguish senders or receivers in the hidden protocol from
other HTTP users contacting the same set of servers.
This enlarges the anonymity set
beyond senders and receivers and provides 
unobservability.

The rest of the paper is organized as follows. In 
Section~\ref{attack}, we define our adversary model. Section \ref{back}
briefly introduces Chaumian mixes and explains why HTTP is
a good choice of cover protocol. Related work is 
examined in Section~\ref{related}. Section~\ref{s2s} describes
a new class of covert channels inside HTTP which allow communication
between servers under the cover of user--generated traffic.
To show how these channels can be put to use in a Chaumian
mix, we present a simple protocol in Section~\ref{mup}.
Unsolved problems and areas for future research are discussed
in Section \ref{disc}.
We conclude with Section~\ref{concl}.
\pagebreak

\section{Threat Model} \label{attack} To mount the timing
and intersection attacks against many employed systems, the
observer only needs to inspect the headers in the layers below
the application layer of the TCP/IP stack at selected points on the
Internet. 

This precisely matches the capabilities of current (legal)
telecommunication surveillance. To cite the
CALEA\footnote{Communications Assistance for Law Enforcement
Act} Implementation Section of the FBI: 
\begin{quotation}
\noindent 
examine[ing] the full packet stream and examine
protocol layers higher than layer 3 would place a high load
on existing network elements in most architectures.
\cite{tiasniff}. 
\end{quotation}
The specification of ``traffic data'' in the EU Convention
on Cybercrime \cite{CoC} indicates that the intended
mandatory surveillance by internet service providers 
is restricted to the lower three layers of the TCP/IP stack.

We grant our adversary the additional ability to
inspect application layer headers. This adversary model
corresponds to an observer who is datamining traffic logs 
for cliques of communicating people. This is 
a realistic and impending threat.

\section{Background}
\label{back}
In Chaumian mixes, nodes relay messages for each other. 
Each node 
has a $(public\,key, private\,key)$ pair. To send a message
along a chain of relaying mixes 
through the mix overlay network, the 
address of the final recipient is attached to the message.
The result is encrypted with the public key of the
last node in the chain. The address of the node is attached
to the result and the process repeated for each node along
the chosen path toward the first. On receipt of a message,
a node decrypts it and --- if it is not the final recipient
itself --- forwards it to the node specified in the decrypted
text.

Later improvements on Chaum's scheme suggest
random delays, various strategies to process and subsequently
dispatch messages ({\em flushing}) \cite{mixmaster-spec,kesdogan,Raymond01},
re--ordering of messages in the pool,  padding the messages
to a fixed size after decryption, and other improvements to
ensure unlinkability. 

Although recently contributed schemes (e.g. MorphMix \cite{RP02-1},
GNUnet's GAP \cite{pet2003-bennett} or Tarzan \cite{tarzan}) require users to 
transport traffic for other users,
many deployed Chaumian mixes and derived systems suffer from
the problem that most users do not --- or perhaps cannot --- run nodes in the systems
themselves. They may be hindered by Network Address Translation \cite{rfc2663},
dynamic --- and therefore unstable --- IP addresses or restrictive firewalling policies.
This greatly weakens the achievable anonymity,
as a passive adversary can observe traffic patterns leading to
and coming from the mix network.

To thwart traffic analysis, we suggest hiding the protocol inside the
well--established HyperText Transfer Protocol (HTTP\cite{rfc2616,rfc2617}). 
According to recent measurements\cite{CAIDA}, 
HTTP accounts for the highest percentage  of
data on the Internet,
only slightly less than FastTrack's \cite{FastTrack} Peer--to--Peer  protocol.
 
Using HTTP as cover traffic brings another
advantage. There is already an
extensive body of research, and several implementations, which aim at
providing some degree of anonymity for HTTP clients
in the presence of various adversaries, see for example
\cite{Crowds}, \cite{Anonymizer}, \cite{LWPA} and \cite{JAP}.
These techniques can be employed to enhance unlinkability.

HTTP is a client--server protocol. At first, this seems to imply
that hidden data can only be forwarded through a chain of
alternating clients and servers, all of which have to be
participants of the hidden network. We will show, however, that
communication between servers is feasable
through standard web--clients which need not be
part of the community using the covert protocol.

\section{Related Work}
\label{related}
The concept of covert channels was introduced by B. Lampson in 1973
\cite{Lampson}.
Covert channels in the network and transport layers of the TCP/IP protocol
were examined by Rowland \cite{Rowland97} and Fisk et al. \cite{Fisk}.
Using HTTP as substrate for other application level protocols
is discussed in RFC~3205 \cite{rfc3205}, where only overt encapsulation
of protocols is considered, naturally.
There are several tools that tunnel protocols through 
HTTP, mostly for circumvention of firewalls, for example,
Lars Brinkhoff's {\tt httptunnel} \cite{Brinkhoff}. These 
tools can be used to disguise any protocol as HTTP traffic,
but the set of entities in which to hide 
(the {\em anonymity set} \cite{Koehn}) consists of just the sender and receiver,
whereas the constructions listed in Section \ref{s2s} use real 
cover traffic, involving unwitting web surfers as cover.
In Infranet \cite{Infranet}, covert channels in HTTP are used to
circumvent web--censorship. Web servers participating in the
Infranet receive hidden requests for censored web pages and
return the pages' content steganographically hidden in harmless images.
Goldberg and Wagner's TAZ and rewebber network \cite{rewebber}
implements anonymous publishing based on HTTP.

In \cite{JAP}, the authors briefly touch on the subject of
unobservability, but conclude that real users would inadvertedly
destroy this property. Surveys such as Raymond's \cite{Raymond01}
mention the concept, but do not point to protocols that provide it.

\section{Server--to--Server Channel\\ through unwitting Clients}
\label{s2s}
In this section, we explain how HTTP servers can communicate
through clients without the consent or knowledge of the user.
This constitutes a new class of covert channel, which transports
data indirectly.
\label{S2S} The main mechanisms inside 
HTTP/HTML that allow such data transmissions are:
\begin{enumerate}
\item Redirects \label{redir}
\item Cookies \label{Cook}
\item Referer\footnote{the typo was in the RFC and stuck.} headers\label{ref}
\item HTML elements \label{elem}
\item ``Active Content'' \label{active}
\end{enumerate}
\noindent
These features can be employed as follows:

\subsubsection{Redirects}
Redirects (RFC 2616  ``303'' messages \cite{rfc2616})
are used to refer the client to another location.\label{RRR}
The location can be the URL of a CGI script, with  optional parameters
in the {\tt QUERY\_STRING} \cite{CGI}.
This allows CGI scripts to send data in said parameters
to other CGI scripts through the browsers of unwitting
web surfers. This channel's capacity is restricted to 
1024~URL--encoded bytes \cite{rfc1738}.

\subsubsection{\bf  Cookies}
Cookies constitute a mechanism to keep state information
on the client side.
To advise the client to keep a {\tt (key,value)} pair 
for further communication, a server sends a {\tt Set-Cookie:}
header in the reply to a request. 
The {\tt value} part is allowed to be up to 4 kilobytes long,
and the standard specifies that a client must be able to store
up to a maximum of 40~cookies per server.
In the server--to--server context, we can use optional
features to transport data between the servers.
The definition of cookies in RFC 2109 \cite{rfc2109}
defines a protocol sub--field {\tt domain} which 
carries information about what group of web servers
the cookie is to be sent to. The RFC states that the
{\tt domain} must contain at least two dots if it ends
in a three--letter Top Level Domain (TLD) and at least three dots
if it ends in a two--letter TLD. There are a many
free Dynamic DNS services online, most of which provide
hostnames in domains with this property, e.g., all hostnames
in the zone administered by {\tt dyndns.org} are in the
same cookie domain. If a CGI script on server
{\tt foo.dyndns.org} sends a cookie of the form
\begin{quotation}
\noindent
{\tt KEY = VALUE; domain = .dyndns.org; Path = /;}
\end{quotation}
to a browser and the browser connects to server \\
{\tt bar.dyndns.org}, then {\tt bar} will get {\tt foo}'s (key,value) pair.
To get the browser to request data objects from {\tt bar.dyndns.org},
the document requested from {\tt foo} could contain one of the
tags mentioned below under ``HTML elements'', or
contains active content that requests data from {\tt bar}
automatically.

\subsubsection{ Referer}
{\tt Referer} headers contain the location of the web page
or script that linked to the presently requested one.
Since the naming of contents can be chosen arbitrarily
by a server --- and forced upon the browser by automatic requests as
described below in subsection \ref{ht} --- 
this is another channel between servers through unwitting browsers.
The length restriction of redirects applies here, too.

\subsubsection{ HTML Elements}\label{ht}
The HyperText Markup Language (HTML) version~4 contains elements that cause 
most browsers to automatically
request given documents from HTTP servers. The following HTML tags 
and attributes have this property:\\
\smallskip
\begin{itemize}
\item{\tt frame src=URL}  Indicates a part of a {\tt frameset}. 
\item{\tt iframe src=URL}  Defines an embedded frame. 
\item{\tt img src=URL}  Defines an inline image. 
\item{\tt script src=URL}  Indicates that JavaScript (see below) functions for
this page should be loaded from {\tt URL}. 
\item{\tt link href=URL}  Indicates out--of--band information for the current page.
\item{\tt object src=URL}  Defines an embedded multi--media object to load. 
\item{\tt applet codebase=URL}  Indicates that Java (see below) classes for
this page should be loaded from {\tt URL}. 
\item{\tt embed src=URL}  Defines an embedded multi--media object to load. 
\item{\tt layer src=URL}  Defines a transparent layer of this page. 
\end{itemize}

If the HTML document is created by a CGI script, the {\tt URL} value in
the tags above can be set to contain the address of another script together
with parameters.\\
The {\tt <META HTTP-EQUIV> } tag/attribute allows embedding of HTTP protocol
header fields in the body of an HTTP message. This is useful
for our purposes, because the header thus embedded in the body
escapes the inspection of our adversary defined in Section \ref{attack}.
Interesting applications in our context are:
\begin{itemize}
\item Redirects (return code~303 \cite{rfc2616}) inside successful
replies (return code~500):\\
{\tt <META HTTP-EQUIV="Refresh"\\
CONTENT="3;URL=http://www.some.org/some.html">}\\
This line of HTML causes the browser to request \\{\tt some.html} from
{\tt www.some.org} after 3~seconds.
\item Setting cookies without a {\tt Set--Cookie} header:\\
{\tt <META HTTP-EQUIV="Set-Cookie" \\
CONTENT="key=value;path=/;domain=.dyndns.org">}\\
This line sets a cookie on the browser, which will be transmitted to
every server in the {\tt dyndns.org} sub--domain to which the browser 
subsequently connects.
\end{itemize}

\subsubsection{ Active Content}
So--called ``Active Content'' is code that is executed on
the client. Currently used languages for
active content are SUN's Java \cite{Java}, Netscape's JavaScript \cite{JavaScript},
Macromedia's Flash \cite{Flash} and
Microsoft's ActiveX \cite{ActiveX}, the
latter being restricted to a single browser, so it
will not be discussed here. 
In Java's design, considerable effort 
was made to make the execution of untrusted code on
the client secure.
Java's security framework inhibits connections to
servers differing from the one which supplied the running
applet, so it cannot be used to transmit data to different servers. 
Of the remaining two languages, we chose
Javascript, because it is more wide--spread and better
documented. Running code on unsuspecting surfer's machines
opens a number of channels of varying bandwidth between scripts 
on servers.  To name two examples:
\begin{itemize}
\item It is trivial to program redirects to CGI
scripts (with parameters) in JavaScript. 
\item A script may construct an invisible {\tt FORM} \cite{HTML4}, fill
the fields with data and send all of it to
a CGI script in the body of a {\tt POST} request
without user interaction.
This channel allows almost arbitrarily large payloads.
\end{itemize}

All the above mechanisms are heavily relied on by authors
of HTML documents and CGI scripts.

\section{The Muted Posthorn --- A Chaumian Mix on Banner Adverts}
\label{mup}
To demonstrate how a anonymous messaging protocol can
use HTTP as cover traffic to achieve unobservability 
against our adversary, we present a simple Chaumian mix.

\subsection{The Setup}
In our variant of Chaum's protocol, the {\em Muted Posthorn},
four (not necessarily disjoint) groups of entities are 
involved:
\begin{description}
\item{\bf The node maintainers} provide CGI scripts on HTTP servers.
The scripts work as mix nodes and so every script has a $(public key,
secret key)$ pair and a pool for messages to be forwarded.
A script is called with the message as the parameter of a POST request.
The scripts work as in Chaum's mix networks, i.e. on receipt of
a message, they decrypt it and look at headers specifying further
processing. In our simple protocol, there are three possible
actions, forwarding the message to another node, storing the
message in a local mailbox with a supplied name (a
128~bit number), and sending
the content of a given mailbox back to the requesting HTTP client.
The outward visible action of the scripts is to return either
an HTML document with JavaScipt code that submits data to another
node, or a short, static HTML document.

\item{\bf The linkers} maintain web pages which all seem to contain 
the same small icon or banner advert. They do this by including an {\tt iframe}
which includes a frameset on one of the nodes.
The frameset consists of a frame with the image and a second, invisible
frame. This frame is created by a node and either
contains the JavaScript code that does the actual transport, or
the short HTML document.

\item{\bf The senders and receivers} use this setup to communicate
encrypted messages. Senders construct messages as in recent mix networks,
e.g. Mixmaster \cite{mixmaster-spec}, but the final delivery
address of a message is always a mailbox on a node, and
special actions must be taken for the first hop in a chain.
A message thus constructed is sent to the first of the nodes 
in the chain by sending a POST request to a script.
Receivers must pull their mailboxes. They do this by
sending encrypted ``send mailbox number N'' requests to the
nodes where they keep mailboxes.

\item{\bf Hapless web surfers} just visit the pages maintained by the
linkers. Their browsers execute the JavaScript code returned
by the node, transfering messages in the process.
\end{description}

\subsection{A first Version}
A simple variant of 
our protocol uses two kinds of messages:
\begin{description}
\item {\tt To:} messages contain encrypted messages to nodes in the network.
\item {\tt Get:} messages request mailboxes from nodes.
\end{description}
Messages are always padded to a fixed length with randomness.
When preparing a message $m_0$ for a sequence of nodes $n_i$, the
sender recursively computes 
$$m_{i+1} = \mathtt{To: } || n_i || E_{n_i}(m_i). $$
where $E_{n}(m)$ encrypts message $m$ for $n$'s public key.
For the last node, the {\tt To}  header is omitted.
The sender submits the encrypted message to the last node
in a POST request.

A receiving node tries to decrypt the message with its secret key.
If decryption succeeds, the resulting text is parsed for headers.

If it is a {\tt Get} message, the node looks up the requested
mailbox. If it exists, the node throws a coin.
On $0$, the content is sent  --- through the
requesting client --- as a message to a random node in the mix network.
The client can extract the message, for example, from its local browser cache.
On $1$, a fixed HTML response is sent to the client.
If the mailbox does not exist, again a coin is tossed, this time
to decide whether to send a randomly chosen message from the pool through
the client or the HTML response.

If it is a {\tt To} message, the address is examined. If it
is a mailbox number, the message is stored in it. If the addressee is a URL,
the message is put in the message pool for further delivery.
Again, a coin throw decides whether a randomly chosen message from the pool
or the fixed HTML response is returned to the client.

Against a passive observer as the adversary defined in
Section \ref{attack}, this protocol provides
unobservability. Senders of messages and requesters of
mailboxes send HTTP GET requests as any harmless client.
Upon receipt of the JavaScript document, they substitute
their own messages for the ones set inside the JavaScript
code, and then let the browser execute the code. An observer
who is restricted to the IP/TCP/HTTP headers thus cannot
distinguish between harmless browsers and senders/receivers.
This increases the anonymity set by the noninvolved
web surfers.

\subsection{DoS attack on the first protocol}
The simple protocol above is susceptible to
a trivial denial of service attack. An adversary can
simply request the frameset from a node repeatedly to drain 
its message pool. 

To defend against this attack, we introduce acknowledgements
for received messages (ACKs) between the nodes. Each message is
kept in the pool and is re--sent until an ACK for the message
is received. ACKs are not sent immediatly, but
are put in the message pool themselves.

An ACK should be tied to the message it acknowledges and 
to the node the message was addressed to, to avoid forged
ACKs and replays. The standard
approach would be to sign ACKs with the node's secret key.
But deploying digital signatures at all would imply that
the nodes know each other's public keys. Experience with
remailers, however, shows that knowledge about such 
a global state of the mix network is hard to achieve.
For this reason we would like to avoid all public key
operations at the nodes, except decryption.

Our suggestion is to send the hash of the decrypted text as
ACK to the previous node (to make them indistinguishable from
other messages, ACKs are padded with randomness to the fixed
message size). The original sender knows all intermediate
messages on the path, since she constructs them layer by layer.
So she can inform every node on the path about what ACK to
expect. She does this by including the ACKs as values of
additional {\tt Ack} headers. The rule for constructing
the next layer is now:
$$m_{i+1} = \mathtt{To: } || n_i || \mathtt{Ack: }|| h(m_i)|| E_{n_i}(m_i). $$

\begin{figure}
 {\fbox {%
    \includegraphics[width=0.95\columnwidth]{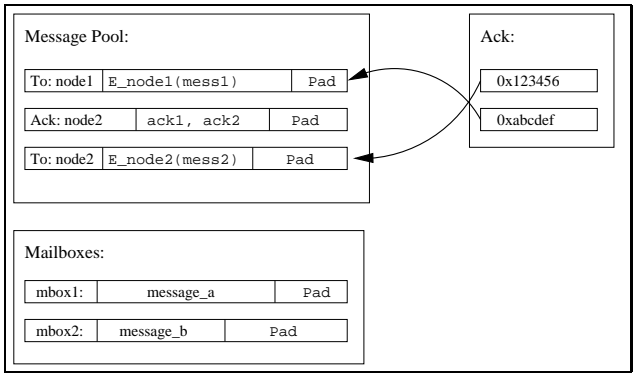}
 }}
 \caption{The internal state of a node: message pool with messages
 and acknowledgements for received messages, ACK table with outstanding
 ACKs and references to messages in the pool, and the mailboxes.}
 \label{fig1}
\end{figure}

A node keeps three tables: the message pool of outgoing messages,
a list of outstanding ACKs and a list of mailboxes (see
figure~\ref{fig1}).

On receipt of a message, a node checks if it is an acknowledgement.
This is done by inspecting the first $|h()|$~bits of the message, where
$h$ is the cryptographic hash function used for ACKs.
The resulting block is checked against the table of outstanding
ACKs. If the block matches, the ACK itself and the message corresponding
to it are removed from the table and the pool, respectively.
Note that the URL of the sending node is transmitted by the client
in the {\tt Referer} header.

When processing {\tt To} messages, the node now creates an entry in
its ACK table with the value of the {\tt Ack} header. 
The node computes the hash of the decrypted message and constructs 
an ACK message for the node that sent the message

\subsection{Properties of the Protocol}
The protocol inherits practical advantages from HTTP. All transactions of
senders, receivers and unwitting web surfers can be performed
through HTTP anonymizing systems such as Anonymizer \cite{Anonymizer},
Crowds \cite{Crowds} or JAP \cite{JAP}.

The protocol's traffic is typically not blocked or modified at firewalls,
and passes though Network Address Translation \cite{rfc2663} 
without problems.

The coin tossing on the nodes makes the auto--submits terminate
after two repetitions, in the mean.
For a fixed message size of four kilobytes, the resulting
traffic for the client is about the same as that for a banner advertisement 
(typically 16~kb). 

\section{Unsolved problems and Directions for future Research}\label{disc}
Although the idea of a mix network with an enlarged anonymity set
seems promising, a number of open problems and possible
enhancements must be discussed.

{\em Do acknowledgements (or lack thereof) introduce new points of attack?}
If a node does not receive an ACK for a message, it will re-send
the message at some later time. The repeating pattern marks
it as being a message as opposed to an ACK or randomness.

{\em User behaviour influences the timing of message delivery.}
This could lead to a Trickle \cite{trickle02} attack.
To reduce this influence, a node could send randomness of 
appropriate size to a randomly chosen node, if a
client connects but the node's
batching strategy does not dispatch a message from the pool.
This would allow reuse of most of the known pooling 
algorithms.

{\em The time a message spends in the mix network before
final delivery is dependent on external factors}, namely
the whims and inclinations
of unknown web surfers, and the willingness of
web--site maintainers (linkers) to place links to the nodes on
their pages. Should all the linker's pages become 
unpopular at some point, communication would stop entirely.

One way around this problem would be to combine the mix network
with an Internet advertising company. The advertisements
(placed in {\tt IFRAME}s) would show ads while at the
same time transporting data between the different servers
of the advertising company. If cookies are used as the channel
of communication, it would not be
noticeably different from what Doubleclick Inc. is doing now \cite{Double}.

{\em Can we achieve unobservability against a
global observer} who inspects complete data payloads, instead of just the headers?
Universal re--encryption \cite{reenc} offers a solution.

In universal re--encryption, a third party (the unwitting
clients, in our case) can change the random factor in a
probabilistic public key encryption, and the following
properties hold:
\begin{enumerate}
\item The third party does not need to know the public key with
which the message is encrypted.
\item For two given encrypted messages, after re--encryption,
an adversary cannot tell which of the outputs corresponds to
which original encryption.
\end{enumerate}
In \cite{reenc}, P. Golle et al. show how universal re--encryption
can be implemented with El--Gamal and a public $(Group,$ $Generator)$
pair. They also show how re--encryption can be extended to
hybrid encryption schemes, where the public key scheme is
used to encrypt a session key and the message itself is encrypted
with a symmetric cipher and the session key.

Unfortunately, JavaScript has no built--in functions for
arithmetic of large numbers nor symmetric ciphers, and 
implemented in JavaScript, they would
be extremely slow. Java, however, offers the {\tt math.BigInteger}
and {\tt SecureRandom} classes necessary for implementing
the re--encryption algorithm. Inconsistent with the security
requirements of Java, standard browser implementations
allow JavaScript to call public methods and variables of Java
objects. JavaScript in turn can be used to submit the re--encrypted
message to the next node, as in the protocol above.

The global observer would see a random--looking message delivered
to the client and another random--looking message from the
client to the next node. 
Because of the properties of El--Gamal (and the 
symmetric cipher in the case of hybrid encryption),
the observer cannot distinguish real messages from randomness.
Because of the properties of universal re--encryption,
she can only guess whether the outgoing message is
a re--encryption of the received one or a completely
new message substituted by the client. Inspection
of the HTTP body does not help to distinguish senders
and receivers from unwitting web surfers. 

Universal re--encryption also remedies the problem
of repeated messages, mentioned above. The
node would re--encrypt the message in the pool before
sending, so that observable messages are always different.

\pagebreak
\section{Summary}
\label{concl}
Privacy is of growing concern for users of the Internet's services.
Existing privacy enhancing technologies can assure anonymity
only if the anonymity set is sufficiently large. In most current
protocols, the size
of the anonyity set is bounded by the number of the active
users of a protocol. 
After defining a reasonable adversary model,
we showed how the anonymity set of a protocol can be
enlarged by having non--participants generate cover traffic.
We presented new covert channels in
the most wide--spread protocol on the Internet, the HyperText Transfer
Protocol, and proceeded to
describe a simple Chaumian mix based on CGI
scripts, in which the anonymity set consist of senders, receivers 
and unknowing participants, thereby enhancing anonymity for
the senders and receivers. We explained remaining problems of our protocol
and suggested areas for future research.

\section{Acknowledgements}
We thank Niels Provos, Marius Aamodt Eriksen, Andrei Serjantov
and Roger Dingledine for helpful suggestions and comments.

\bibliographystyle{abbrv}
\bibstyle{acm}

\end{document}